\documentclass[aps,prl,twocolumn,groupedaddress,draft,showpacs,intlimits,amsmath,amssymb,floats]{revtex4}
\newcommand{\bk}{\textbf{k}}

\newcommand{\ket}[1]{|#1\rangle}
\newcommand{\braket}[2]{\langle #1|#2\rangle}

\usepackage[final]{graphicx}
\usepackage{color}

\begin{document}


\title{Surface adatom conductance filtering in scanning tunneling 
spectroscopy of Co-doped BaFe$_{2}$As$_2$ iron pnictide superconductors}
%

\author{K. Koepernik$^{1}$}
\author{S. Johnston$^{1}$}
\author{E. van Heumen$^{2}$}
\author{Y. Huang$^{2}$}
\author{J. Kaas$^2$}
\author{J. B. Goedkoop$^2$}
\author{M. S. Golden$^{2}$}
\author{Jeroen van den Brink$^{1}$}

\affiliation{$^1$
Institute for Theoretical Solid State Physics, IFW Dresden, Helmholtzstrasse 20, 01069 Dresden, Germany}
\affiliation{$^2$ van der Waals - Zeeman Institute, University of Amsterdam, Sciencepark 904, 1098 XH Amsterdam, the Netherlands}

\begin{abstract}
We establish in a combination of {\it ab initio} theory and experiments that the 
tunneling process in scanning tunneling microscopy/spectroscopy on the A-122 iron 
pnictide superconductors - in this case BaFe$_{2-x}$Co$_x$As$_2$ - involve a strong 
adatom filtering of the differential conductance from the near-$E_F$ Fe3d states, 
which in turn originates from the top-most sub-surface Fe layer of the crystal.  
The calculations show that the dominance of surface Ba-related tunneling pathways 
leaves fingerprints found in the experimental differential conductance data, 
including large particle-hole asymmetry and an energy-dependent contrast inversion. 
\end{abstract}

\date{\today}
\pacs{74.70.Xa,73.40.Gk,73.20.-r}
\maketitle
The discovery of high-temperature superconductivity in 
the iron pnictides has prompted a tremendous effort directed towards understanding 
the nature of superconductivity in these systems. Theory indicates 
unconventional superconductivity but with a pairing symmetry that can be very 
sensitive to electronic and structural details of the system \cite{PeterReview}. 
Experimentally, 
surface sensitive probes of the electronic structure such as angle-resolved photoemission 
spectroscopy (ARPES) \cite{TerashimaPNAS,KoitzschPRL2009,WiPNAS2011,EvHPRL2011} 
and scanning tunneling microscopy/spectroscopy (STM/STS) \cite{Hoffman,MasseePRB2009,
MasseeEPL2010,ZhangPRB2010,ChuangScience2010,NascimentoPRL2009,YinPRL2009}, play 
a major role in the search for a microscopic understanding of the superconductivity 
in these systems. 
In this context, the alkaline-earth-metal AFe$_2$As$_2$ (A-122, A = Ca, Ba, Sr) 
iron pnicties have been widely studied due to the availability of high-quality 
crystals, and the possibility to both n- and p-type dope these systems. 
It is therefore essential to definitively 
establish the properties of A-122 surfaces.  Accessing the underlying physics 
representative of the bulk relies on a quantitative understanding of surface-related 
matrix elements coupled to potential surface-related  changes in electronic/structural 
properties and this is exactly what we aim to do here.

Perhaps the most prominent theater for the surface/bulk debate is the 
case of BaFe$_{2-x}$Co$_x$As$_2$ (Ba-122) \cite{Hoffman}.  
Due to the strong covalent bonding within the AsFe$_2$As block, the 
surface of cleaved Ba-122 is expected to be either an exposed FeAs 
layer  or a half layer of Ba atoms. Although there are indications for the 
former \cite{NascimentoPRL2009}, there is a much larger body of evidence for the latter  
\cite{EvHPRL2011,MasseePRB2009,MasseeEPL2010,ZhangPRB2010,YinPRL2009}. 
Electrostatics would dictate a half Ba layer as the crystal termination. Both the 
strong temperature dependence of the reconstructed surface and 
the large variety of topographic images observed by STM \cite{MasseePRB2009,Hoffman} 
are most easily understood in terms of relatively loosely bound Ba 
atoms (which could be considered adatoms) constituting the termination 
layer. Density functional theory (DFT) data point to a Ba termination 
with a $\sqrt{2}$$\times$$\sqrt{2}$ or $2\times$1 
reconstruction as energetically favorable \cite{GaoPRB2010}. Recently, a combined 
LEED/ARPES/DFT study showed that the experimental LEED I/V curves could only result 
from a reconstructed Ba surface and furthermore that the associated surface-related 
bands overlapping with the bulk states have a strong impact on the breadth of spectral 
features observed in ARPES for this system  \cite{EvHPRL2011}.

A quantitative understanding of tunneling processes in complex, layered crystals 
is important, as is illustrated in the cuprate high T$_c$ systems, where asymmetry 
between electron removal and addition in STS data is seen as a sign of 
strong correlation effects \cite{HanaguriNature2004}, as well being attributed 
to the effect of the 
specific tunneling pathway from the CuO$_2$ plane LDOS through the SrO and BiO 
layers to the STM tip \cite{NieminenPRL2009}. 
Given the marked $2\times1$ and $\sqrt{2}$$\times$$\sqrt{2}$ surface reconstructions 
observed in the LEED and STM experiments on the A-122 pnictide family \cite{Hoffman}, 
it is important to uncover the impact of the real surface structure on spectroscopic 
tunneling measurements themselves.

We tackle this issue here by theoretically establishing the precise 
role of the Ba surface in STS of Co-doped Ba-122 superconductors, 
showing that the surface Ba atoms strongly filter the tunneling current, 
leading to marked particle-hole asymmetry in the tunneling conductance 
$dI/dV$ and bias-dependent contrast inversion in $dI/dV$ maps. A direct 
comparison with experiment explicitly confirms these key predictions.
%

\begin{figure}
 \begin{center}
 \includegraphics[width=\columnwidth]{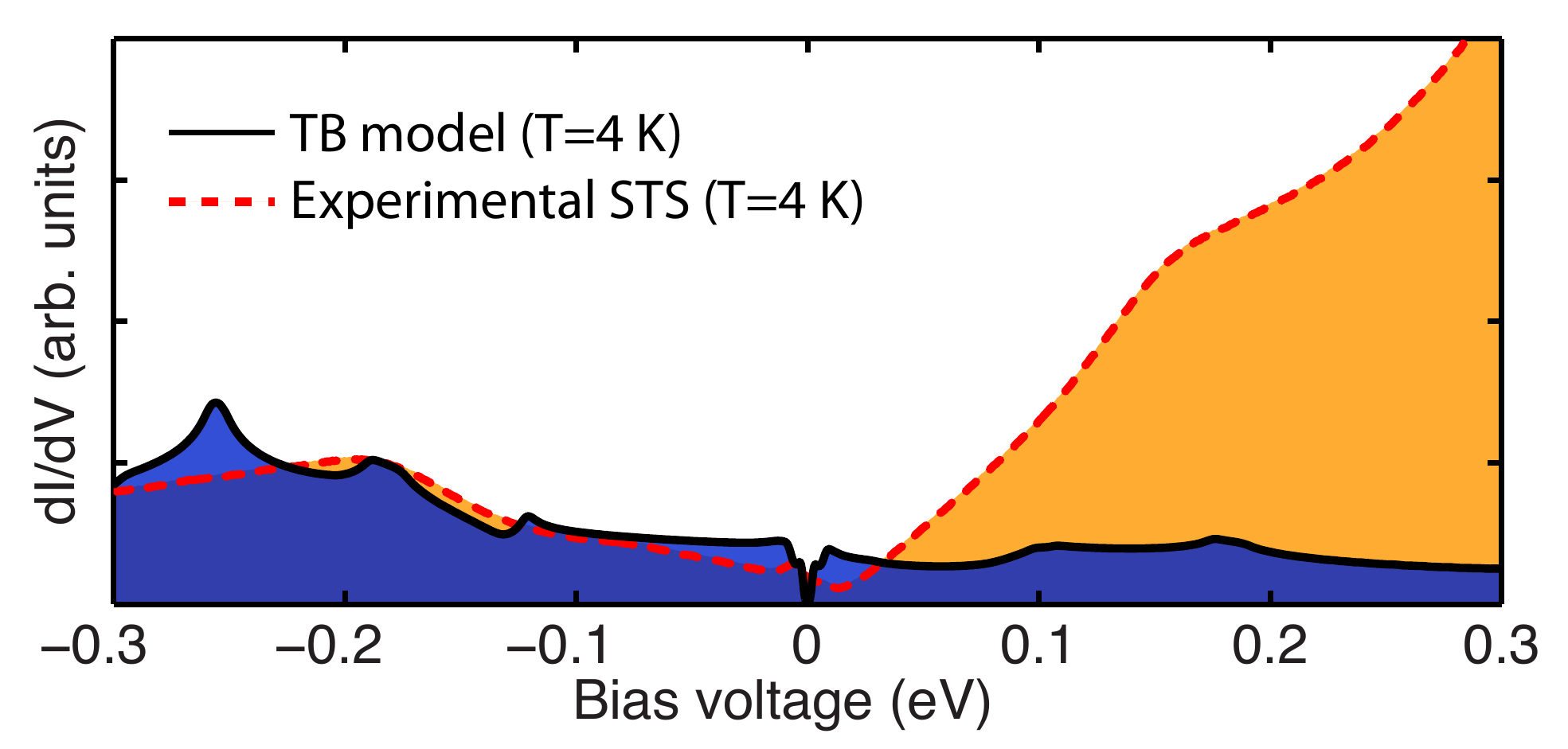}
 \end{center}
 \caption{\label{Fig:TB}
 (Color online)
 Comparison of calculated and measured $dI/dV$ spectra in Co-doped 
 Ba-122.  The model spectrum (black) was obtained from a five-band tight 
 binding model for Ba-122 including the effects of interband impurity scattering 
 within the superconductor.  The experimentally measured spectrum (red/dashed 
 line, $V_{bias} = -300$ meV, $I_{set} = 50$ pA) was recorded at 4 K on an 
 overdoped sample (x = 0.22, T$_c$ = 12 K).
 }
\end{figure}

We start by examining the tunneling conductance 
in a simplified Fe band-based model, incorporating the 
five-band tight-binding model of Graser {\it et al.} \cite{Graser} 
and comparing the results to the experimental data.
This model was obtained by fitting to DFT calculations and so an   
additional two-fold renormalization of the bands is included to account for 
electronic correlations.  
As a next step, superconductivity is introduced at the BCS level, 
including interband impurity scattering. Since the interest here is 
in $dI/dV$ over energy scales almost 100 times that of $\Delta_{sc}$, 
the precision of the theoretical description of the superconductivity 
is non-critical \cite{Supplement}.  Fig. \ref{Fig:TB} compares the model $dI/dV$ to   
measured spectra obtained on an overdoped sample (x = 0.22 i.e. 11\% doping, T$_c$ = 12 K).  
On the occupied side - negative sample bias - the gross structure is 
reproduced by the model, and one might assign the broad peak in the STS 
trace at $\sim200$ meV to features in the Fe-based DOS.
However, the unoccupied states raise doubts as to whether such a simple 
interpretation is tenable, as the measured $dI/dV$ exhibits a large 
particle-hole asymmetry with a much larger signal at positive bias voltage, 
a characteristic wholly lacking in the data of the 
tight-binding model. This discrepancy cannot be 
removed by including additional self-energy effects as they would only  
serve to further broaden the spectral features. It is clear that the theory 
needs to leave the confines of the five Fe-band basis, and 
the remainder of this letter is devoted precisely to this, showing that recognizing 
the filtering effect of the surface Ba states on the tunnel current removes 
the discrepancy so visible in Fig. \ref{Fig:TB}.  

To model the tunneling process correctly we have performed DFT calculations to simulate tunneling 
into various finite slab geometries of 8.5\% ($x = 0.17$) Co-doped Ba-122.
These calculations use the full potential local orbital code (FPLO),
\cite{FPLO}, in the scalar-relativistic and local density approximations (LDA) \cite{pw92}.  
The choice of functional is not critical, since we are considering the
non-magnetic state and fixed atomic geometries. See \cite{Supplement} for
further computational details.

The Ba-122 surface is modeled using a slab structure with a 2$\times$1 
reconstructed, half Ba termination layer with the atomic positions fixed to 
LEED data \cite{EvHPRL2011}.  Since STM/STS experiments are often performed 
with Pt/Ir tips, we model the tip with a single Pt atom (henceforth referred 
to as the `tip'), positioned at different locations above the sample. 
The tunneling conductance is calculated from a weighted overlap between 
the slab wave functions and the localized
6$s$ and 5$d$ states of the tip atom: 

\begin{equation}\label{Eq:dIdV}
\frac{dI}{dV}\propto \sum_{n,\bk,i} \delta(\mu - \epsilon_{i}^t)
\delta(\mu + eV - \epsilon_{\bk,n}) |\braket{\Psi_{\bk,n}}{\chi_i^t}|^2
\end{equation}
where $\ket{\chi_i^t}$ is the discrete tip state $i$  
with energy $\epsilon^t_i$, and $\mu$
is the Fermi level of the slab. The $\delta$ function
representing the tip density of states (DOS) is replaced with a
Lorentzian broadened expression, mimicking band formation in a real 
macroscopic tip. Here we used $\gamma = 1$ eV for all tip states, 
noting that the exact $\gamma$ value has no qualitative influence 
on the results, and merely results in a scaling of the magnitude 
of the resulting $\frac{dI}{dV}$ curves. Eq. (\ref{Eq:dIdV}) is 
essentially a version of the Todorov-Pendry approximation \cite{Pendry91} 
as has been used in a wave function context \cite{Kobayashi95} and a 
Green's function context \cite {Suominen11}. It goes beyond the 
commonly used Tersoff-Hamann model \cite{tersoff85}, which is derived for  
$s$-wave tips only.

The matrix element in Eq. (\ref{Eq:dIdV}) in its full form is
$M(\bk,n)=\braket{\Psi_{\bk,n}}{V_T\mid \chi_i^t}$,
with $V_T$ the potential at the tip. However, our tests have shown 
that for a nearly spherical potential
the effect of including $V_T$ also only results in a global rescale of the
$\frac{dI}{dV}$ curves, and thus the form shown in Eq. (\ref{Eq:dIdV}) is
sufficient.

\begin{figure}[b]
 \includegraphics[width=\columnwidth]{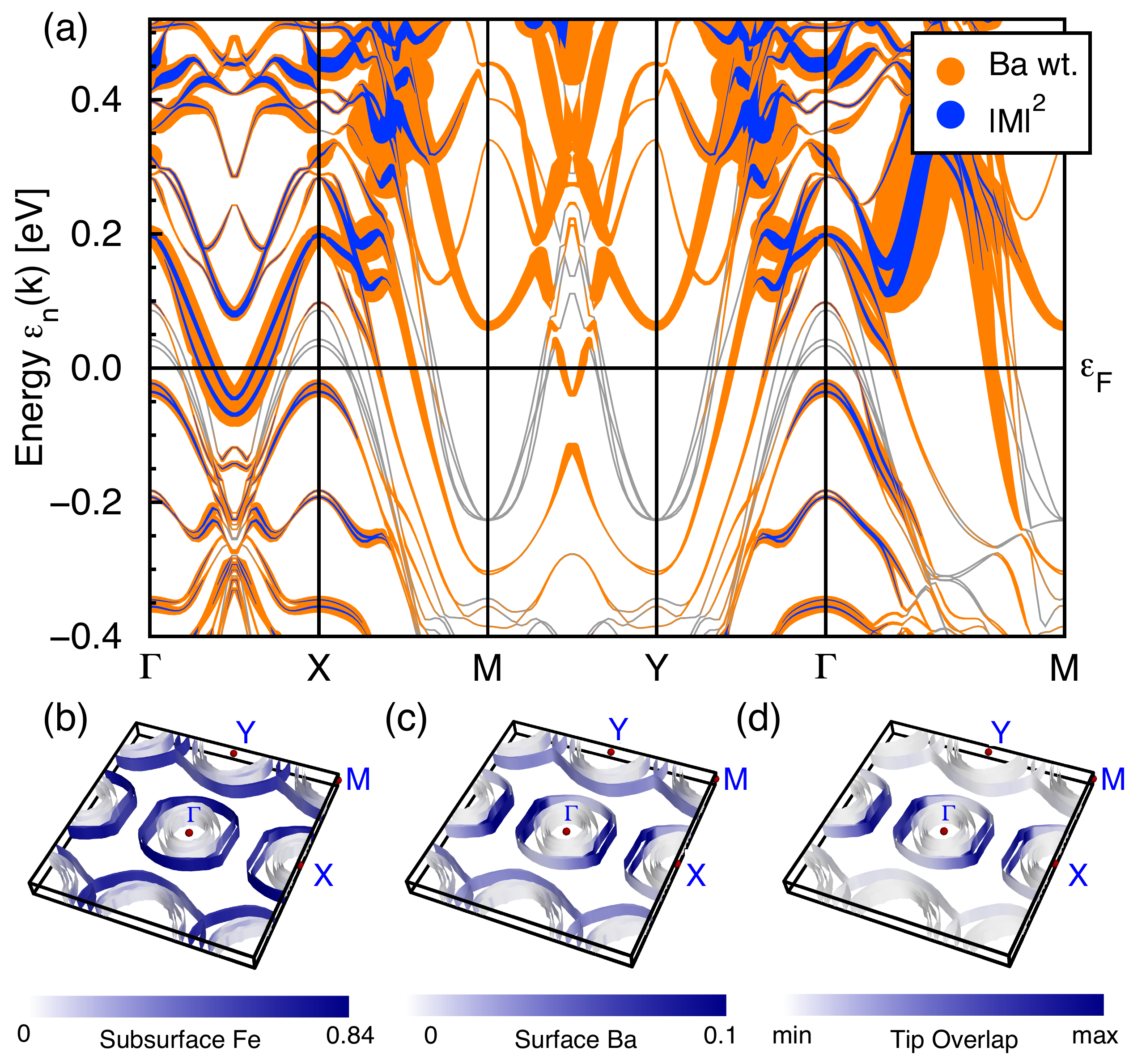}
 \caption{
 \label{Fig:BaWgt} (Color online) (a) The band structure of the 2$\times$1 
 surface reconstructed Ba-122 slab. The Ba weight of the band is coded by the thickness of the 
 orange (light) shading. The weight of the blue (dark) shading indicates the 
 tip-sample tunneling matrix element. (b) the character of the closest-to-surface 
 Fe and (c) surface Ba orbitals, respectively, projected onto the Fermi surface. 
 (d) A similar Fermi surface projection of the tip-sample tunneling matrix element 
 illustrating a strong correlation with the Ba projection shown in (c).
 }
\end{figure}

We first consider the matrix element for tunneling between the tip and the
sample $M(\bk,n) = \braket{\Psi_{\bk,n}}{\chi_r}$. 
Fig. \ref{Fig:BaWgt}(a) plots the band structure for the $2\times$1 reconstructed slab, 
with numerous surface bands from reconstruction-induced back-folding and atomic distortions 
in the outermost As-Fe$_2$-As block \cite{EvHPRL2011}.  
The thickness of the orange (light) shading codes for the weight associated with the surface-Ba 
character of the bands. Most of the Ba-dominated bands lie above $E_F$, except for one band 
with Ba-$5d$ character dipping below $E_F$ between $\Gamma$ and $X$.  
The magnitude of the tip-sample matrix element $|M(\bk,n)|^2$, coded by the weight of the 
blue (dark) shading, is also shown in Fig. \ref{Fig:BaWgt}(a).   
Comparing the two shadings, it is clear that the largest tunneling matrix elements 
coincide with the presence of bands with strong surface Ba character.
To illustrate this important finding, panels (b)-(d) of Fig. \ref{Fig:BaWgt} 
show the Fermi surface of the slab.  In panel (b) the blue shading 
codes the sub-surface Fe 3d character, in (c) the Ba surface character and in (d) 
the tunneling matrix element $|M(\bk,n)|^2$. Panel (c) reflects the surface reconstruction, 
in that the Ba projection clearly breaks C4 symmetry. 
The essential point here is that panels (c) and (d) strongly resemble each other, 
meaning regions of the Fermi surface with the largest surface-Ba weight are
directly correlated with the largest tunneling matrix element.
In contrast, the weight of the Fe atoms is isotropic around the Fermi surfaces 
(Fig. \ref{Fig:BaWgt}[b]) and does not correlate with the suppressed tunneling 
matrix element along the $\Gamma$-$Y$ direction evident in panel (d). 
At low bias voltage, the primary tunneling pathway is in fact one from the uppermost 
Fe layer of the terminating As-Fe$_2$-As block to the tip, and this runs through the 
Ba surface atoms, involving the $\Gamma$-$X$ region of the Brillouin zone.   
It is through this filter that STS-experiments see the superconducting states 
of these materials.

\begin{figure}[t]
 \begin{center}
  \includegraphics[width=1.0\columnwidth]{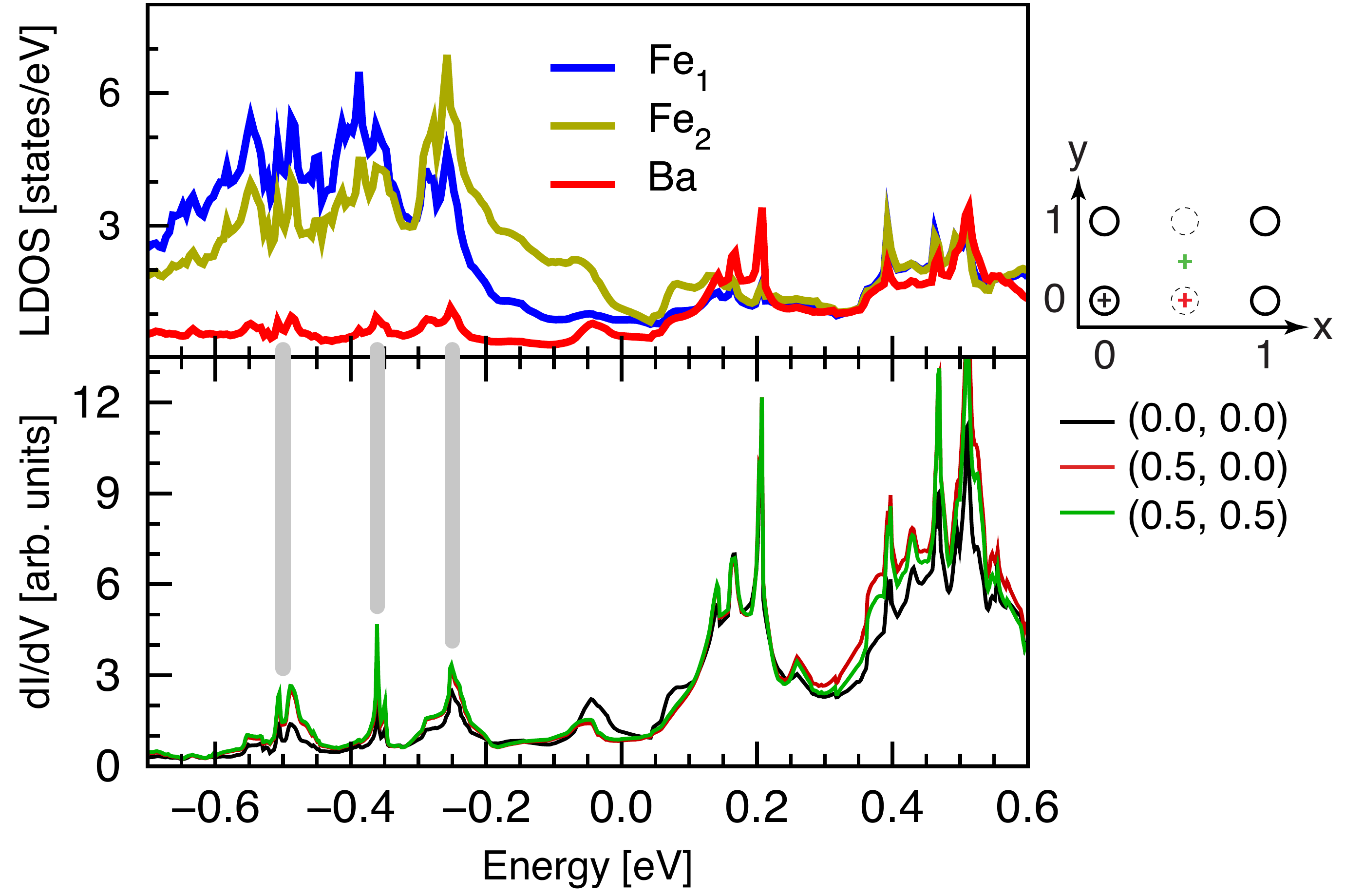}
 \end{center}
 \caption{
 \label{Fig:dIdV} (Color online) The tunneling conductance $dI/dV$ for the 2$\times$1 
 reconstructed surface slab, calculated using Eq. (\ref{Eq:dIdV}), for various 
 tip positions relative to the crystal lattice. The coordinate scheme is sketched 
 in the top right of the figure and $x=0$, $y=0$ corresponds to a tip located directly 
 above a Ba atom in the 2$\times$1 termination layer. Upper panel: projection of the 
 density of states onto the $3d$-orbitals from two representative sub-surface Fe 
 (blue, yellow) atoms and the $6s/5d$-orbitals from the Ba (red) atom of the surface layer. 
 Here all $dI/dV$ spectra have been normalized over the energy range [0,300] meV.}
\end{figure}

Having established the clear correlation between the tunneling matrix element and 
Ba character of the bands, we now turn to the calculated 
tunneling conductance $dI/dV$ itself. 
In Fig. \ref{Fig:dIdV} the calculated $dI/dV$ spectra obtained for various 
tip locations above the $2\times$1 terminated A-122 (A=Ba) surface are shown. 
In simple models $dI/dV$ is expected to be a measure of the local DOS of the sample 
at the tip position and so the partial DOS, projected on the surface Ba atoms 
and two representative sub-surface Fe atoms is plotted in the upper panel.
Inspecting the lower panel of the figure, the first observation is that the 
calculated $dI/dV$ spectra exhibit a large bias asymmetry, with tunneling into 
unoccupied states of the sample clearly being favored, regardless of the exact 
tip position. Thus, consideration of $dI/dV$ rather than LDOS now brings the 
theory into line with the experimental data (Fig. \ref{Fig:TB}) in terms of 
tunneling asymmetry.  Furthermore, the important intermediary role of the 
adatom states leads to the observation that the main features of the 
spectra for both electron removal and addition correlate well with the 
structure in the Ba projected partial DOS. 
If it were not for the tunneling matrix element in Eq. (\ref{Eq:dIdV}) the 
simulated $dI/dV$ curve would resemble the LDOS of the Fe atoms, showing - 
with respect to the experimental data - the `wrong'  asymmetry with the 
conductance larger on the occupied, negative sample bias side. 

Looking at the three different traces in the lower panel of the figure, 
one sees that the the main features of the tunneling spectrum are 
insensitive to the lateral position of the tip position with respect 
to these high symmetry positions within the $2\times$1 adatom lattice. 
Experimentally, the measured STS spectra are indeed uncorrelated with the 
topographic details within the $2\times$1 and $\sqrt{2}$$\times$$\sqrt{2}$ 
surface reconstructions \cite{MasseePRB2009}, which lends credibility 
to the relevance of both the model and the simulation parameters used 
here.

In the simulations, for experimentally 
relevant tip$\leftrightarrow$sample separations, the filtering of the 
tunneling current by the adatoms, especially the asymmetry of the 
$dI/dV$ spectra, is also robust for 
$\sqrt{2}$$\times$$\sqrt{2}$ surfaces. However, the peak structure in 
the  calculated $dI/dV$ curves on the occupied side depends on the 
assumed locations of the sub-surface Fe and As atoms: their 
positions determine the fine details of the surface-related band splittings  
(see online supplementary material).

In the computer, variation of the tip-sample separation goes without risk of a
tip crash, and our simulations show that dependence of the conductance curves
on the lateral tip position increases if the tip is moved closer to the
surface.  For the data shown in Fig. \ref{Fig:dIdV}, a vertical
tip$\leftrightarrow$Ba adatom distance of $14\AA$ was used, motivated by the
experimental situation. 
At tip sample distances well below those used typically in experiment (i.e.
with the tip pushed to within $3\AA$ of the surface), our simulations show that
the filtering effects through the Ba adatoms are reduced. 

\begin{figure}[t]
 \includegraphics[width=\columnwidth]{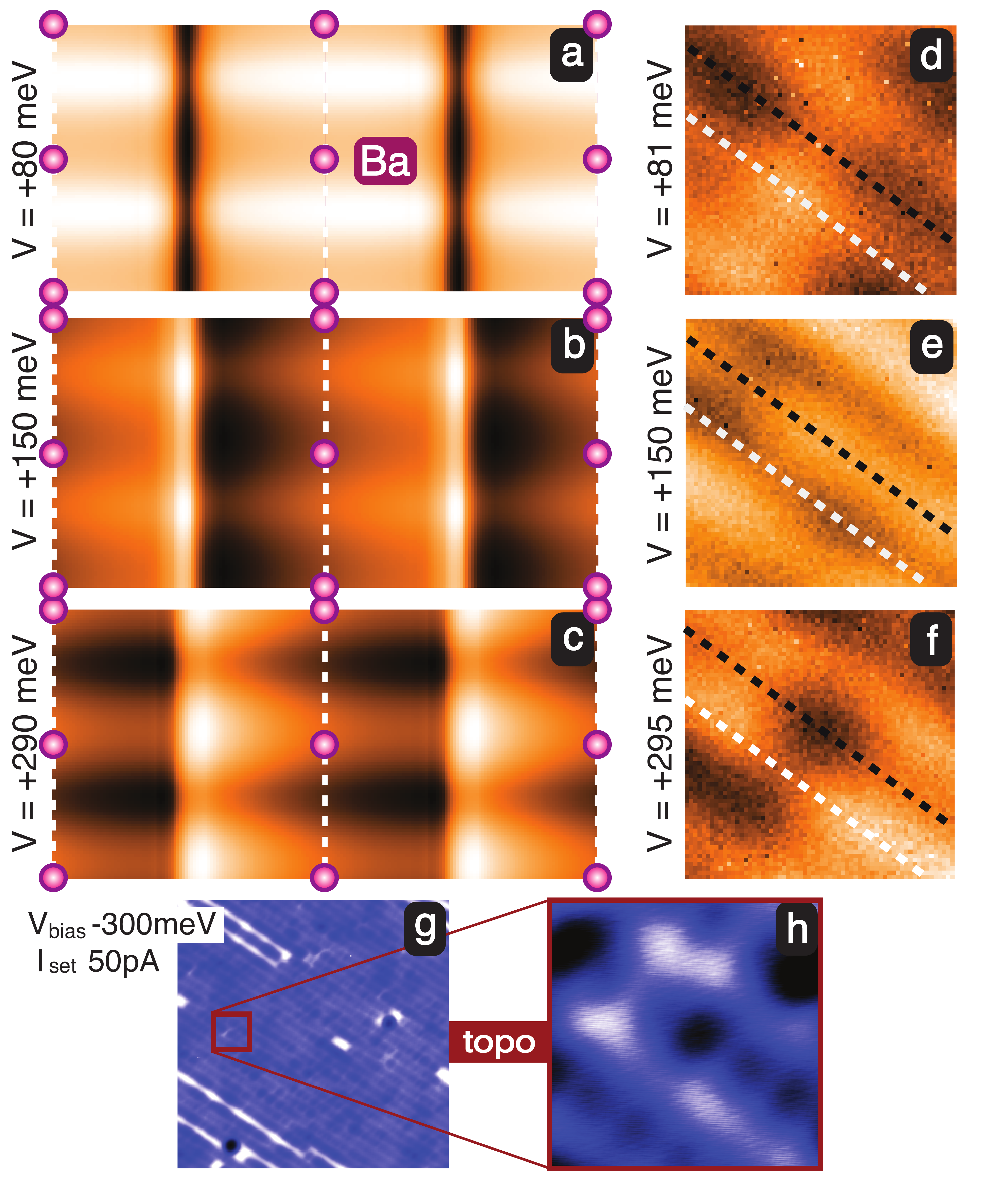}
 \caption{\label{Fig:Maps} (Colour online) (a)-(c) DFT-calculated $dI/dV$ maps
 for $2\times$1 Ba-122 for the energies indicated.  A clear contrast-switch is
 observed between 80 and 295 meV positive bias, in which the high conductance
 shifts from being on to between the Ba.  (d)-(f) Experimental $dI/dV$ maps for
 a $20\times 20$ $\AA$ field of view from a 15\% Co-doped Ba-122 crystal ($x =
 0.3$, $T_c = 12$ K) at the positive bias voltages indicated.  The 
 black (white) dashed lines
 are guides indicating the (missing) Ba $2\times$1 rows.  
 (g) A topographic image covering
 $200\times 200$ $\AA$. The orange square indicates the FOV of panels (d)-(f),
 the topographic data for which is shown in panel (h). 
 }
\end{figure}

Going further than the three different lateral tip positions in Fig.
\ref{Fig:dIdV}, we now simulate tunneling conductance maps, constructed by
varying the tip atom's lateral position at a fixed distance from the surface of
the slab (14 $\AA$, as before).  
This generates a set of $dI/dV$ spectra as a function of tip position, which in
this case are normalized such that the integral from $[0,300]$ meV is fixed. To
create the conductance map for a particular bias voltage V, a slice through the
simulated data-cube at this bias is then plotted as a function of $r$.   

Applying this procedure to the simulated data for the $2\times$1-reconstructed
surface produces the maps shown in panels (a) - (c) of Fig. \ref{Fig:Maps}, for
$V = 80$, $150$, and $295$ meV, respectively. At $V=80$ meV, the  $2\times$1
reconstruction is clearly seen in the calculated maps, appearing - for example
- as a dark stripe joining the locations of the missing Ba atoms.  
Interestingly, as the bias voltage is increased to 150 meV (panel [b]) a clear
contrast reversal takes place.  This behavior in the model is
confirmed in our measured tunneling conductance maps, shown in panels (d)-(f)
of Fig. \ref{Fig:Maps}. These were measured with a 5 meV modulation on the bias
voltage and had set-up parameters $V_{set} = -300$ meV and $I_{set} = 50$ pA,
with normalization of the $dI/dV$ spectra to the area of the electron addition
spectral weight up to 300 meV.  The (black) white dashed lines in Fig.
\ref{Fig:Maps}(d)-(f) are guides, marking the Ba adatom peaks (valleys) 
observed in the topography (panel [g]), of which a zoom is presented in Fig.
\ref{Fig:Maps}(h).    
The measured $dI/dV$ maps show a similar relative contrast
reversal at the same positive bias
values.  The peaks (valleys) in the conductance map at 
80 meV (panel [d]) reverse roles at 150 meV, appearing    
instead as valleys (peaks) (panel [e]).  If one traces along the 
occupied Ba row under the white dashed line in panel [e] at 150meV bias, the
experiment indicates higher dI/dV between the positions of the Ba atoms (the
latter visible in the topographic scan shown in panel [h] as bright features)
and the corresponding theory map (panel [b]) shows the same behaviour. Moving
up to a bias around 290meV in panel (c),
a second type of contrast change 
occurs along the occupied Ba row where the occupied (unoccupied) Ba positions 
become brigher (darker).   
This change is also reflected in the experimental data (panel [f]).  
However, at 295 meV the agreement 
between theory and experiment is not perfect: 
theory predicts the highest intensity to occur in between the Ba rows 
whole in the experimental data the highest intensity occurs on the Ba rows. 
Nevertheless,  
the experimental confirmation of the contrast changes at lower bias voltage 
predicted by our DFT calculations closes the loop as regards the 
validity and precision of our theoretical description showing a heavily 
Ba-filtered tunneling current in STS of Co-doped Ba-122 superconductors.  

In summary, our joint theoretical and experimental investigation has uncovered
that the alkaline earth adatom termination layer in STS experiments on A-122 iron
pnictide superconductors acts as a tunneling filter. Bands with strong Ba5d
character in the Ba-122 system - which originate in the $2\times$1
reconstruction - are the door-keeper to a sub-set of the Fe3d related states
from the uppermost Fe layer of the sub-surface As-Fe$_2$-As block.    Our model
naturally explains the observed strong particle-hole asymmetry observed in the
STS spectra, which cannot be found in the underlying Fe 3d partial DOS.
We expect very similar adatom-filtering effects to impact the tunneling current
in other alkaline-earth A-122 systems, but given the sensitivity to the surface
reconstruction, experimental data regarding the real surface structure - such
as those from LEED - are required before the different aspects can be
successfully unravelled as we have done here.  

\acknowledgments
The authors acknowledge F. Massee, A. Kemper and
A. V. Chubukov for useful discussions. This work was supported
by the Foundation for Fundamental Research on Matter (FOM) 
and the VENI program (both part of of the Dutch NWO) and the EU
(FP7/2007-2013, no. 226716).

\end{document}